\documentclass[a4paper,fleqn,usenatbib,useAMS]{mnras}

\usepackage{newtxtext,newtxmath}

\usepackage[T1]{fontenc}
\usepackage{ae,aecompl}

\usepackage{graphicx}	
\usepackage{amsmath}	
\usepackage{amssymb}	
\usepackage{bm}	

\title[Thermal responses of a coronal loop]{Thermal responses in a coronal loop maintained by wave heating mechanisms}

\author[Takuma Matsumoto]{Takuma Matsumoto \thanks{E-mail:takuma.matsumoto@nagoya-u.jp}\\
Institute for Space-Earth Environmental Research, Nagoya University, Chikusa-ku, Furo-Cho, 464-8601 Nagoya, Japan}

\date{Accepted XXX. Received YYY; in original form ZZZ}

\pubyear{2017}

\begin{document}
\label{firstpage}
\pagerange{\pageref{firstpage}--\pageref{lastpage}}
\maketitle

\begin{abstract}
A full 3-dimensional compressible magnetohydrodynamic (MHD) simulation is conducted to investigate the thermal responses of a coronal loop to the dynamic dissipation processes of MHD waves.
When the foot points of the loop are randomly and continuously forced, the MHD waves become excited and propagate upward.
Then, 
1-MK temperature
corona is produced naturally as the wave energy dissipates.
The excited wave packets become non-linear just above the magnetic canopy, and the wave energy 
cascades into smaller spatial scales.
Moreover, collisions between counter-propagating Alfv\'{e}n wave packets increase the heating rate, resulting in 
impulsive temperature increases. 
Our model demonstrates that the heating events in the wave-heated loops can be nanoflare-like 
in the sense that they are spatially localised and temporally intermittent.
\end{abstract}

\begin{keywords}
Sun: chromosphere -- Sun: corona -- MHD -- turbulence -- waves.
\end{keywords}



\section{Introduction}
It is widely accepted that the couplings between magnetic fields and surface convections are
the primary energy source of coronal heating \citep[e.g.][]{2006SoPh..234...41K}.
Energy injection time scales are conventionally used to classify the coronal heating mechanisms into two categories.
Energy injections with time scales longer than the Alfv\'{e}n crossing time result in stressing models \citep{1981ApJ...246..331S},
whereas those with shorter time scales result in wave models \citep{1947MNRAS.107..211A}.
However, it has not yet been determined which coronal heating model is actually in effect in the solar atmosphere.
In this paper, we will focus on the wave models and examine the possibility of explaining the properties observed in the
active region loops.

The dynamic properties of coronal heating mechanisms have been studied intensively under the assumption of reduced magnetohydrodynamics (RMHD).
RMHD frameworks are considered applicable to plasma contained in tokamaks that is low-beta and incompressible 
and has a high aspect ratio \citep{1976PhFl...19..134S}, which are the properties of plasma in the solar wind and the solar corona.
RMHD frameworks suggest that the coronal plasma above the open-field regions is heated dynamically via 
interactions between counter-propagating Alfv\'{e}n waves \citep{1999ApJ...523L..93M,2003ApJ...597.1097D,2007ApJ...662..669V}.
Recently, these ideas have also been adopted for closed coronal loops, including those comprising the effects of the chromosphere
\citep{2011ApJ...736....3V,2012A&A...538A..70V}. However, it must be remembered that RMHD assumptions are only marginally valid 
for the chromosphere.
In addition, these models are unable to describe the thermal responses caused by the dynamic dissipation of Alfv\'{e}n waves 
because they ignore thermal processes such as radiation and conduction.

In the present study, the thermal responses of coronal plasma to wave-heating mechanisms are investigated under fully compressible 3D MHD equations, which cannot be considered within RMHD frameworks.
Recently, numerous studies have investigated the properties of MHD waves, such as energy flux and energy ratios, among different wave modes 
\citep{2003ApJ...599..626B,2005ApJ...631.1270H,2011ApJ...727...17F,2015ApJ...799....6M}. However, these studies did not focus on thermal responses to wave dissipation.
More realistic models that include radiative processes and thermal conduction \citep{2015ApJ...811..106H,2017ApJ...834...10R}
can describe the thermal balance in the corona properly.
These models are self-consistent in that the cooling caused by radiation and conduction is balanced
by the dynamic heating processes that are described by the non-linear MHD equations.
Both \cite{2015ApJ...811..106H} and \cite{2017ApJ...834...10R} suggest that the braiding of the field lines
causes coronal heating, and there are only a few self-consistent multi-dimensional models sustained by waves 
\citep{2016MNRAS.463..502M}.

From an observational point of view, 
localised and episodic heating events, or nanoflares, \citep{1983ApJ...264..642P}
are considered to be required to explain the properties of coronal loops
\citep{2002ApJ...579L..41W}.
The overpressure loops observed in EUV \citep{2001ApJ...550.1036A} and the delays between loops in 171 \AA and 195 \AA 
\citep{2003ApJ...593.1164W} suggest that the coronal loops are not in thermal equilibrium state but in dynamically evolving state.
The fact that the lifetimes of observed loops are longer than expected can be explained by loops that consist
of a number of thin strands heated sequentially.
These observations require the intermittent and spatially-localised nature of nanoflares.

The purpose of this study is to investigate the nanoflares in wave-heated loops and determine their effects on the thermal properties
of the loops using fully compressible 3D MHD simulations.
To accomplish this, we utilize a simplified model composed of a single flux tube that is easy to analyse.
Using 1D MHD simulations based on a simplified model, \cite{2004ApJ...601L.107M} suggested that the fast and slow shocks converted from
nonlinear Alfv\'{e}n waves can maintain a hot coronal loop. Moreover, \cite{2010ApJ...712..494A} investigated 
the regime under which Alfv\'{e}n wave heating produces hot and stable corona.
\cite{2016MNRAS.463..502M} extended these studies via 2D simulations, revealing that the contributions of a turbulent cascade 
across the loop axis become comparable or larger than those of the MHD shocks.
Using 3D RMHD simulations, \cite{2011ApJ...736....3V} showed that heating processes became impulsive in wave-heated coronal loops, 
although the thermal responses to these heating events have not yet been determined. For this reason, we investigate the thermal responses of a wave-heated coronal loop.

\section{Models and Assumptions}

In the present study, we attempt to mimic a single coronal loop using fully compressible 3D MHD simulations.
Considering the active region loops, we assumed slender loops 100 Mm in length and 3 Mm ($\equiv W$) in width.
For simplicity, half of a straight loop in a rectangular domain ($x\in[0,50]$ Mm, $y\in[-1.5,1.5]$ Mm, 
and $z\in[-1.5,1.5]$ Mm) was considered, and the effects of curvature were ignored. 
The compressible MHD equations, including the effects of radiation and conduction, were solved based on Cartesian geometry:

\begin{equation}
 \frac{\upartial \rho}{\upartial t}+\nabla \cdot (\rho \mathbfit{v})=0,
\end{equation}

\begin{equation}
 \frac{\upartial \rho \mathbfit{v}}{\upartial t} + \nabla \cdot \left(
 p+\frac{B^2}{2} + \rho \mathbfit{v}\mathbfit{v}-\mathbfit{B}\mathbfit{B}
 \right)=\rho \mathbfit{g} + \mathbfit{F}_{\rm ex},
\end{equation}

\begin{equation}
 \frac{\upartial \mathbfit{B}}{\upartial t} + \nabla \cdot \left( \mathbfit{v}\mathbfit{B} - \mathbfit{B}\mathbfit{v}\right)=0,
\end{equation}

\begin{eqnarray}
 \frac{\upartial {\cal E}}{\upartial t} &+& \nabla\cdot\left[ \left( {\cal E}+p+\frac{B^2}{2}\right)\mathbfit{v}
 -(\mathbfit{B}\cdot\mathbfit{v} )\mathbfit{B}\right] \nonumber \\
 &=&\rho \mathbfit{v} \cdot \mathbfit{g}+\nabla \cdot \left( \kappa \nabla T \right) + Q_{\rm rad} + \mathbfit{F}_{\rm ex} \cdot \mathbfit{v},
\end{eqnarray}

\begin{equation}
 {\cal E}=\frac{p}{\gamma-1}+\frac{\rho v^2}{2}+ \frac{B^2}{2},
\end{equation}
where $\rho$, $\mathbfit{v}$, $p$, $\mathbfit{B}$, ${\cal E}$, and $T$ are the mass density, fluid velocity, 
gas pressure, magnetic field normalized against $\sqrt{4\upi}$, total energy density, and temperature, respectively.
$\kappa$, $Q_{\rm rad}$, and $\mathbfit{g}$ represent the Spitzer-type thermal conductivity tensor, 
radiative cooling function, and gravitational acceleration of a half circle loop, respectively.
The thermal conduction in our model is effective only along the magnetic field.
When the temperature was less than $4\times 10^4$K or the density was greater than $4.9\times10^{-17}$ g cm$^{-3}$, 
we used the empirical cooling rate of $9\times10^9 \rho$ erg cm$^{-3}$ s$^{-1}$ 
as a rough approximation of optically thick cooling \citep{1989ApJ...346.1010A}. 
Otherwise, we applied optically thin cooling \citep[e.g.][]{1990A&AS...82..229L}.
The functional forms of $\kappa$, $Q_{\rm rad}$, the equation of state, and $\mathbfit{g}$ 
can be found in \cite{2014MNRAS.440..971M,2016MNRAS.463..502M}.

\begin{figure}
 \begin{center}
  \includegraphics[width=8.5cm]{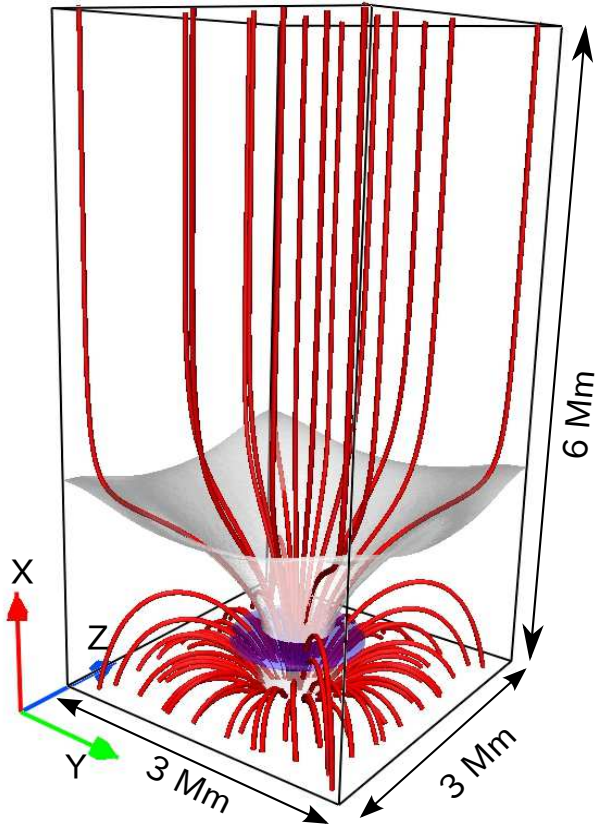} 
 \end{center}
\caption{
Initial potential magnetic field structures below $x=6$ Mm. 
Red tubes represent magnetic field lines, and grey surface indicates an isosurface at which plasma beta is equal to 1.
Blue surface corresponds to region in which external force was applied.
}\label{fig:init_b}
\end{figure}

The initial conditions for our simulations were similar to those used in \cite{2016MNRAS.463..502M}.
A hydrostatic equilibrium with a temperature of 10$^4$ K and with a bottom density of 10$^{-7}$ g cm$^{-3}$
, which is the typical photospheric density,
was assumed at $x\le$ 15 Mm, while the density distribution above $x=$ 15 Mm was inversely proportional to 
the square of the height but had the same temperature.
Thus, the initial atmosphere was not in a dynamic equilibrium state, and materials started to descend just after 
our simulations began.
We set a non-equilibrium state to avoid severe CFL condition due to low density at the initial phase.
The initial magnetic field was assumed to be a potential field that was extrapolated 
from a given photospheric magnetic field distribution, $B_x(x=0,y,z)$:
\begin{eqnarray}
 B_x(x=0,y,z) \equiv B_0 \exp \left[ -\frac{y^2+z^2}{w_B^2}\right] + \Delta B,
\end{eqnarray}
where $B_0=$ 1,800 G and $w_B=$ 500 km. $\Delta B$ was a free parameter, and its value in this study was set
such that the field strength at the top of the loop was 10 G.
In Fig. \ref{fig:init_b}, the initial configuration of the magnetic field lines are shown.
The solid red tubes represent the initial magnetic field lines, and the shaded grey surface 
indicates an isosurface where the plasma beta ($2p/B^2$) is equal to 1.

Ghost cells with velocities of zero were placed below the lower boundary.
The density of the ghost cells was determined using a linear extrapolation of the logarithmic density, and symmetrical boundary condition was applied to the temperature.
In addition, a line-tied condition was applied to $B_x$ at the lower boundary.
The other components of the magnetic field were extrapolated at the third order of accuracy.
At the top of the loop, symmetrical boundary conditions were applied to all of the variables.

To simulate the motions of a flux tube jostled by granular cells, external forces were introduced:
\begin{eqnarray}
 \mathbfit{F}_{\rm ex} &=& \mathbfit{f}(t) J_0\left(\frac{r \alpha_0}{r_0}\right) H(r_0-r) \exp \left[-\frac{\left( x-x_f\right)^2}{2 w_f^2}\right], \\
 r&=&\sqrt{y^2+z^2},
\end{eqnarray}
where $J_0$ and $H$ are 0th Bessel function of the first kind and the Heaviside function, respectively, and
$r_0$, $x_f$, and $w_f$ were set to values of 1 Mm, 0.4 Mm, and 0.1 Mm, respectively.
$\alpha_0$ is the first zero point of $J_0$.
In Fig \ref{fig:init_b}, the blue shaded area indicates the region in which external forces were applied.
The temporal evolution of the external forces was modelled by $\mathbfit{f}(t)$, which had a total power 
of 2.3$\times$10$^{-4}$ g$^2$ cm$^{-4}$ s$^{-4}$ and a white noise spectrum with a finite range of
$\nu \in [2.5\times 10^{-4},2\times 10^{-2}]$ Hz.
The amplitudes of the external forces were fine-tuned so that the root mean square of the horizontal velocity 
became 6 km s$^{-1}$ at $x=2$ Mm, which was consistent with a previous 2D simulation \citep{2016MNRAS.463..502M}.
The resultant photospheric ($x<0.2$ Mm) velocity was approximately 1.6 km s$^{-1}$, which was within the reasonable velocity range of 
0.5 to 5 km s$^{-1}$ in the intergranular lanes \citep{1996ApJ...463..365B}.
The topology of the excited velocity field can be modified using the different spatial functions for the external force, 
although we used translational force as the first step.

The numerical scheme adopted in our simulation was the ADER-WENO scheme \citep{2009JCoPh.228.5040B,2009JCoPh.228.2480B}. 
Third orders of accuracy in both space and time were achieved by using WENO spatial reconstruction and ADER temporal integration.
To preserve the positivity, we replaced WENO-reconstructed slopes with minmod slopes 
around the strong shocks and the temperature transition regions.
For the numerical flux, we adopted the HLL approximated Riemann solver from \cite{1991JCoPh..92..273E}.
In addition, the solenoidal constraint, $\nabla \cdot \mathbfit{B}=0$, 
was guaranteed be limited to within round-off errors for this scheme.

The number of grids in each direction, $({\rm N_x,N_y,N_z})$, was $(1024,64,64)$.
The horizontal grid lengths were uniform, measuring 47 km in both the $y$ and $z$ directions. However,  
the vertical grid length was non-uniform, increasing from 25 km at the bottom to 93 km at the top.

\section{Results}
\begin{figure}
 \begin{center}
  \includegraphics[width=8.5cm]{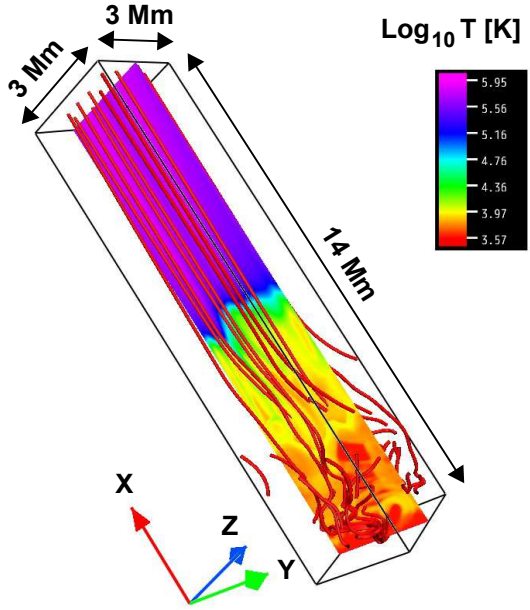} 
 \end{center}
\caption{
Snapshot of temperature and magnetic field structures after the system reached quasi-steady state ($t=72$ min).
Red lines represent magnetic field lines, and the
temperature distribution is shown in the middle plane.
}\label{fig:te_steady}
\end{figure}

As time passed in the simulation, the model of the atmosphere separated into two 
distinct layers: a cool, dense layer similar to the photosphere and the chromosphere, and a hot, rarefied layer similar to the corona.
Driven by the continuous forces at the foot points of the loop, 
the MHD wave packets propagated upward to transport their wave energy.
Because the external force in our model was translational, not only Alfv\'{e}nic waves but also 
highly compressible waves were generated simultaneously at the foot point. 
In addition to the generation at the foot point, nonlinear mode conversion would occur at the magnetic canopy to
excite the compressible waves.
More than 99-\% of the energy of the upward-moving wave packets was lost below the transition region while
the remaining energy was consumed by coronal heating.
Collisions between the wave packets and the transition region revealed vertical motions similar to spicules
\citep[e.g.][]{1999ApJ...514..493K,2017ApJ...848...38I}.
The spicules have 15-40 km s$^{-1}$ velocity, 6-8 Mm height, and 4,000-7,000 K temperature, 
because of which they could be classified as type II spicules \citep{2007PASJ...59S.655D,2012ApJ...750...16Z,2012ApJ...759...18P}.
Because the resulting coronal layer was maintained for well over an hour, the corona was considered to exist in
a quasi-steady state.
In this state, statistical values, including the mean temperature and density averaged over the coronal region,
fluctuated about their temporal averages.
Fig. \ref{fig:te_steady} shows the temperature and magnetic field configuration at $t=72$ min.
The solid red tubes indicate the magnetic field lines, and the coloured plane represents 
the temperature distribution.

\begin{figure}
 \begin{center}
  \includegraphics[width=8.7cm]{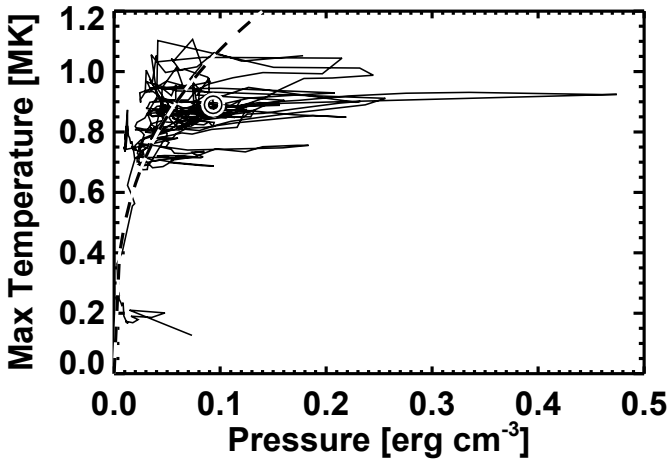} 
 \end{center}
\caption{
Evolution of thermal properties of the coronal loop in the plane of maximum temperature 
and pressure averaged over the transition region (solid line).
Dashed curve represents RTV scaling law ($T=1400(PL)^{1/3}$) with half-loop length, $L$, of 45 Mm, and circle shows the maximum temperature averaged over time.
}\label{fig:rtv}
\end{figure}

The mean maximum temperature, transition region pressure, and half-loop length averaged over $t\in[50,80]$ min 
were 0.9 MK, 0.9 erg cm$^{-3}$, and 45 Mm, respectively.
The temporal average was calculated over the same duration throughout the rest of this study unless otherwise stated.
The solid lines in Fig. \ref{fig:rtv} show the temporal evolution of the system in the $P-T$ plane from $t\in[50,80]$ min, 
where $P$ is the pressure in the transition region and $T$ is the maximum temperature.
In addition, the circle in the plane shows the maximum temperature averaged over the same duration, and the dashed line indicates the series of thermal equilibrium states that were estimated based on RTV theory with a half-loop length of 45 Mm \citep{1978ApJ...220..643R}.
Our model reproduced a loop that was $\sim$2 times denser than the loop predicted by RTV theory, 
which indicates either that the system was heated non-uniformly \citep{2001ApJ...550.1036A}
or that our system was not in a state of thermal equilibrium.

\begin{figure}
 \begin{center}
  \includegraphics[width=8.7cm]{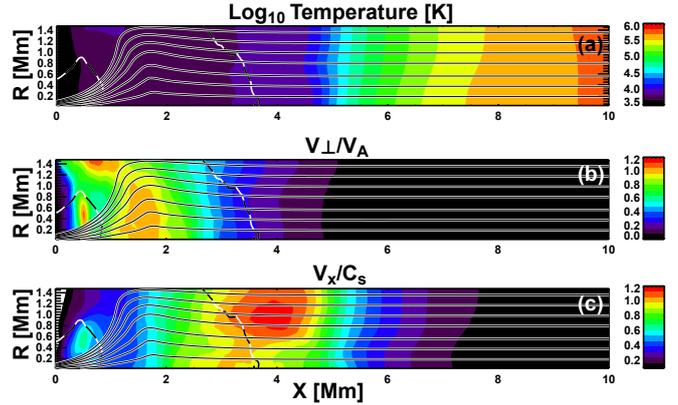} 
 \end{center}
\caption{
Model atmospheres averaged over time about loop axis; (a) flow speed over Alfv\'{e}n speed; (b) flow speed over sound speed; (c) temperature, where solid white lines enveloped by black lines represent 
magnetic field lines, and dashed black-and-white lines indicate contours at which plasma beta is 1.
}\label{fig:average_2d}
\end{figure}

The atmospheric structures averaged over time about the loop axis are shown in Fig. \ref{fig:average_2d}.
Fig. \ref{fig:average_2d}-(a) shows the temperature distribution as a function of $x$ and the radius of the flux tube $(R)$.
The cool chromosphere and the hot corona were separated by the transition region at approximately $x=$ 5 Mm.
The solid black lines enveloped by the white lines represent the magnetic field lines extended from the photospheric magnetic concentrations.
The flux tube expanded super-radially at heights below $\sim$ 1.5 Mm, while the magnetic field was almost uniform above that height,
because neighbouring flux tubes appeared periodically in the $y$ and $z$ directions.
In this model, we refer to a height of $\sim$ 1.5 Mm as the magnetic canopy, 
although several different definitions of this height have been given \citep{2000SoPh..194...29Z}.
The dashed black-and-white lines show the contours at which the value of the plasma beta is equal to 1.
The region near the foot points and above $\sim$ 4 Mm had plasma beta values lower than 1.

The fluctuating horizontal flow speed ($\sqrt{v_{\rm y}^2+v_{\rm z}^2}$) normalized by the local Alfv\'{e}n speed 
provided a measure of the nonlinearity of the transverse MHD waves propagating in the $x$ direction. This value varied with height because of the magnetic, stratified atmosphere.
The nonlinearity increased with height, achieving a peak value greater than 1 
just above the magnetic canopy, and it decreased with height to less than 0.1 in the corona 
(Fig. \ref{fig:average_2d}-b).
This trend can be explained qualitatively by the linear theory of the Alfv\'{e}n wave, which suggests that the nonlinearity of such waves
is proportional to $\rho^{1/4}B^{-1}$.
The nonlinearity increased with height under the magnetic canopy 
because the effects of the decreasing field strength outweighed those of the density.
Above the magnetic canopy, the field strength stopped decreasing, while the density decreased continuously, 
resulting in a decrease in nonlinearity.
This trend depended on the expansion profile of the flux tube, which was affected by the filling factor of the magnetic concentration and the width of the flux tube.

The Sonic Mach number, defined as the vertical flow speed ($|v_{\rm x}|$) normalized by local sound speed, provided a measure of plasma compressibility.
The spatial distribution of the sonic Mach number is shown in Fig.  \ref{fig:average_2d}-c.
Below the transition region, the sonic Mach number approached a value of 1, whereas it decreased to lower than 0.1 in the corona.
In the linear theory of acoustic waves, the wave amplitude normalized by local sound speed, or the sonic Mach number, 
is proportional to $\rho^{-1/2}T^{-3/4}$ above the magnetic canopy.
Because the magnitude of the temperature inhomogeneity was small below the transition region, 
the sonic Mach number increased with height due to the decreasing density.
The sudden temperature increase in the transition region pushed the sonic Mach number below 0.2 in the corona.

The total energy flux at the coronal bottom 
($x\sim7$ Mm)
was $\sim$ 10$^5$ erg cm$^{-2}$ s$^{-1}$, which was lower than 
the energy flux required above the quiet regions \citep{1977ARA&A..15..363W}.
Approximately $5\times10^{-4}$ of the photospheric energy flux reached the corona.
The energy reduction is a product of the transmission coefficient 
and the areal ratio of the open field region to the entire domain in the photosphere.
Because the areal ratio was about 6\% in our model, the transmission coefficient was estimated to be less than 1\%.



\begin{figure}
 \begin{center}
  \includegraphics[width=8.7cm]{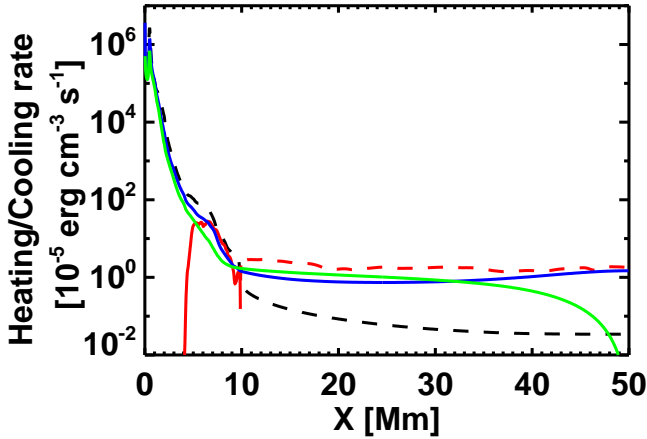} 
 \end{center}
\caption{
Heating and cooling rates averaged over time across the cross-section of the loop.
Black, red, blue, and green lines represent the heating (solid) and cooling (dashed) rates caused by 
radiation, thermal conduction, viscosity, and resistivity, respectively.
}\label{fig:ebalance}
\end{figure}

In our model, the 
1-MK temperature 
corona was maintained against radiative and conductive cooling via heating processes.
Fig. \ref{fig:ebalance} shows the heating (solid lines) and cooling (dashed lines) rates averaged 
over time across the loop cross-section.
Black, red, blue, and green represent the heating/cooling rate of the radiation, thermal conduction, viscosity, and resistivity, respectively.
Radiative cooling was the dominant cooling process below the coronal bottom, 
while thermal conduction losses were dominant in the corona.
The magnitudes of the viscous and resistive heating rates were nearly the same over the entire loop except at its top.
At the loop top, the resistive heating rate was zero because the magnetic fluctuations were also zero due to the symmetrical boundary conditions.
The scale height of the total heating rate fluctuated at approximately 1 Mm below $x=$10 Mm and increased from 10 Mm to $\sim$ 100 Mm above $x=$10 Mm.
Although dissipation occurred at the grid scale because of numerical diffusion, 
the dynamic processes that transported the kinetic and magnetic energy from the system-size 
to the scales of $\ga$ 10 grids would be described properly.
The numerical heating rate was estimated based on the dissipative flux of the HLL Riemann solver, because the explicit dissipation terms were not included in the basic equations.
We defined the decreases in magnetic and kinetic energy caused by numerical diffusion as resistive and viscous heating, 
although they did not fully correspond to the physical resistive and viscous processes.

\begin{figure}
 \begin{center}
  \includegraphics[width=8.5cm]{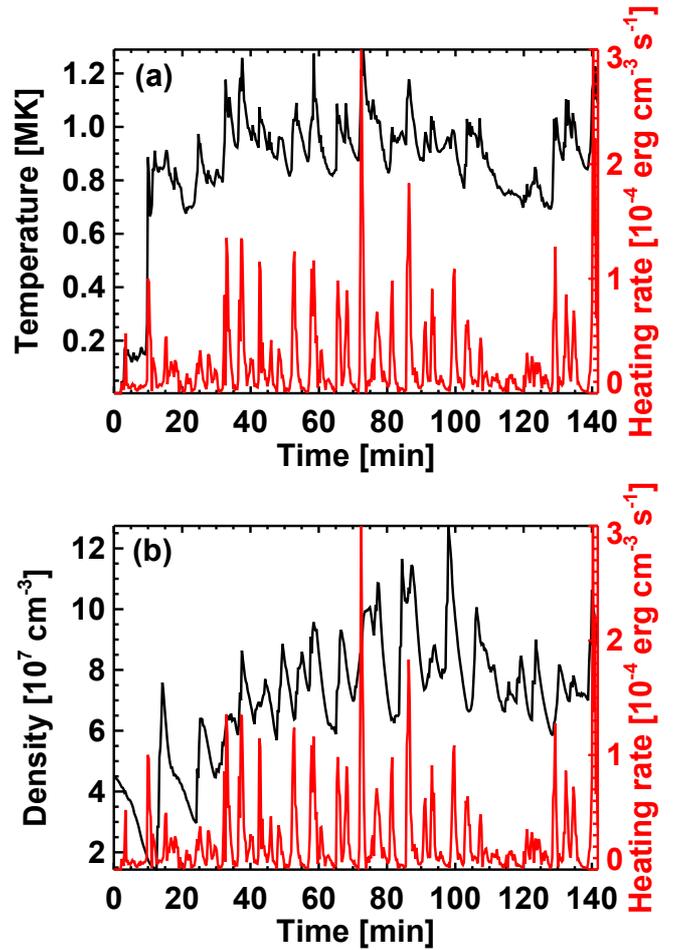} 
 \end{center}
\caption{
Time series of (a) maximum temperature of the loop and (b) density at the loop top.
The mean heating rate (red solid line) is depicted in the secondary axes of both figures.
}\label{fig:te_evolve}
\end{figure}

The maximum temperature measured for the entire loop system revealed a series of impulsive peaks that exhibited sudden increases and gradual decreases between 0.8 MK and 1.3 MK (Fig. \ref{fig:te_evolve} a).
Timeline of density at the loop top is also shown in figure \ref{fig:te_evolve}b.
The timelines indicate that both temperature and density nearly achieved the quasi-steady state after $t\sim50$ min.
The timeline of the temperature-peaks matched well with that of the heating rate at the loop top.
The heating rate of the peaks was typically ten times larger than the averaged values.
The initial increase in temperature at around $t\sim10$ min was caused by the shock waves resulting from the falling of the materials that were not in dynamic equilibrium at $t=0$.
Because the initial dynamics were unrealistic, we ignored the dynamics at $t<20$ min.

\begin{figure}
 \begin{center}
  \includegraphics[width=8.7cm]{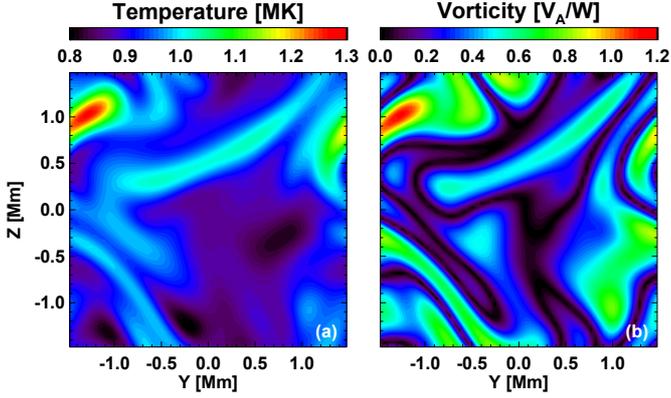} 
 \end{center}
\caption{
 Spatial distributions of temperature (a) and absolute vorticity (b) across loop axis at its apex for $t=58.5$ min.
 Vorticity is normalized by local Alfv\'{e}n speed divided by loop width.
}\label{fig:te_om_snap}
\end{figure}

The spatial distribution of the heating rate was localised in the region that exhibited a strong vorticity.
Fig. \ref{fig:te_om_snap}-a represents a snapshot of the temperature distribution across the loop top at $t=58.5$ min.
Only a small portion of plasma in the loop was heated to temperatures higher than 1 MK.
The high-temperature regions were almost always associated with the high vorticity regions, 
while the converse was not necessarily true (Fig. \ref{fig:te_om_snap}-b).
The widths of the vorticity sheets were approximately 0.2 Mm in the current simulation, which was most likely a result of the numerical resolution.

The energy cascade toward smaller spatial scales was more significant in the chromosphere ($x\in[2,5]$ Mm) than in the 
corona ($x>20$ Mm).
The power indices of the power spectral density (PSD) of the horizontal velocity were 
1.5$\sim$ 2.0 in the chromosphere and $>$4.0 in the corona. These values must be dependent on the resolution.
We obtained these power indices by fitting the PSDs between [180, 1000] km.
The PSD, $P(x,k\perp)$, was defined such that
\begin{eqnarray}
 \langle v_{\perp}^2 \rangle = \int P(x,k_\perp) dk_\perp,
\end{eqnarray}
where $\langle v_\perp^2 \rangle$ is the mean square of the horizontal velocity.

\begin{figure}
 \begin{center}
  \includegraphics[width=8.7cm]{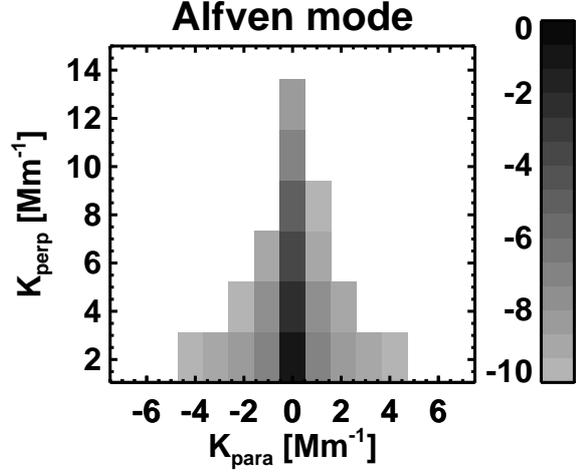} 
 \end{center}
\caption{
Velocity power spectral density of the Alfv\'{e}n mode at the top of the loop.
Horizontal and vertical axes indicate wave numbers in the $x$ direction and direction across loop, respectively, while colour indicates logarithmic power spectral density normalized by total power.
}\label{fig:powspec_alfven}
\end{figure}

It was determined that the Alfv\'{e}n mode was the dominant wave mode in the corona ($x\in[38,44]$).
As all MHD fluctuations can be expanded as linear combinations of the MHD eigen vectors ,
we can specify the dominant MHD mode if we know the eigen vectors.
Assuming the background field is uniform, we decomposed all MHD fluctuations into different MHD modes 
\citep{2003MNRAS.345..325C}.
The contribution of the Alfv\'{e}n mode to the total energy of the fluctuation was approximately 95-\%, 
the contribution of the slow mode was approximately 5-\%,and that of the fast mode was less than 1-\%.
Fig. \ref{fig:powspec_alfven} represents the estimated velocity PSD of the Alfv\'{e}n mode, $P(k_{\rm x},k_\perp)$.
The PSD is defined such that
\begin{eqnarray}
 \langle v_{\rm alfv\'{e}n}^2 \rangle = \iint 2 \upi k_\perp P(k_{\rm x},k_\perp) dk_{\rm x} dk_\perp,
\end{eqnarray}
where $\langle v_{\rm alfv\'{e}n}^2 \rangle$ is the mean square of the velocity amplitude in the Alfv\'{e}n mode.
The PSD revealed that the anisotropic energy cascade in the direction perpendicular to the background magnetic field was similar to those found in previous studies \citep[e.g.,] []{1983JPlPh..29..525S}.
Because MHD mode decomposition in highly inhomogeneous atmosphere like the chromosphere is not trivial we applied this
analysis only in the corona. 
Because kinetic energy in the $x$ direction ($\rho v_x^2/2$), which
can be a measure of compressible wave energy,
is comparable with the remaining kinetic energy, the contribution of the compressible waves can be 
comparable with the Alfv\'{e}nic waves in the chromosphere.

\begin{figure}
 \begin{center}
  \includegraphics[width=8.7cm]{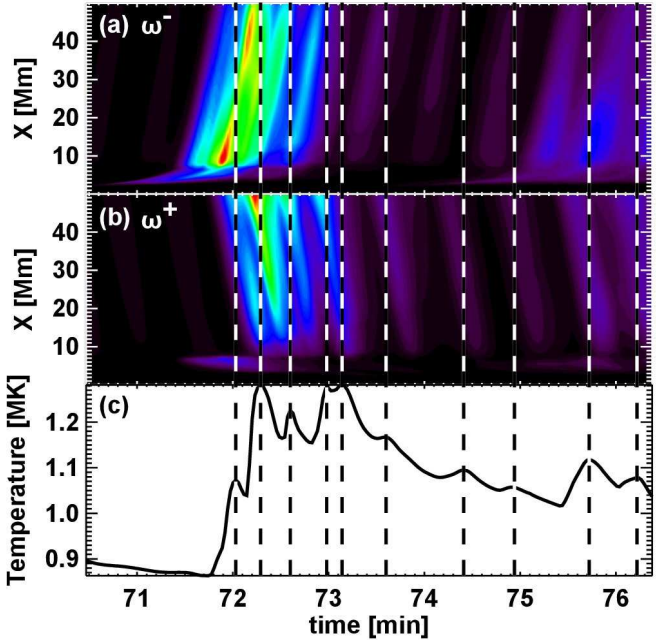} 
 \end{center}
\caption{
Relationship between collisions of Alfv\'{e}n waves and impulsive heating events. Panels (a) and (b) show time-distance diagrams of the square of upward and downward Els\"{a}sser vorticities, while panel (c) shows time evolution of the maximum temperature around temperature peak at $t=72$ min 
in Fig. \ref{fig:te_evolve}.
At each temperature peak, dashed black-and-white lines are over-plotted.
}\label{fig:omg_transport}
\end{figure}

It was determined that the timeline of the heating events that occurred at around $t=72$ min matched well with that of the collisions between counter-propagating Alfv\'{e}n waves.
The temporal evolution of the maximum temperature is shown in Fig. \ref{fig:omg_transport}-c 
in the same way as in Fig. \ref{fig:te_evolve} but with more sampling points.
In addition to the global temperature peak for this period, a number of small temperature peaks were observed, which are indicated by vertical dashed lines.
To investigate whether these peaks were related to the collisions of Alfv\'{e}n waves, Els\"{a}sser vorticity was used to trace the propagating signature of the Alfv\'{e}n waves.
The Els\"{a}sser vorticity is defined as
\begin{eqnarray}
  \bmath{\omega}^\pm \equiv \nabla \times (\mathbfit{v}\pm\mathbfit{B}/\sqrt{\rho}),
\end{eqnarray}
where the plus/minus superscripts represent the vorticity of downwardly/upwardly propagating Alfv\'{e}n waves 
in our configuration.
Fig. \ref{fig:omg_transport}-a,b shows the time-distance diagram of the square of $\omega_x^{\pm}$.
Based on this figure, it was determined that each temperature peak was associated with a collision between the Alfv\'{e}n waves.

\begin{figure}
 \begin{center}
  \includegraphics[width=8.7cm]{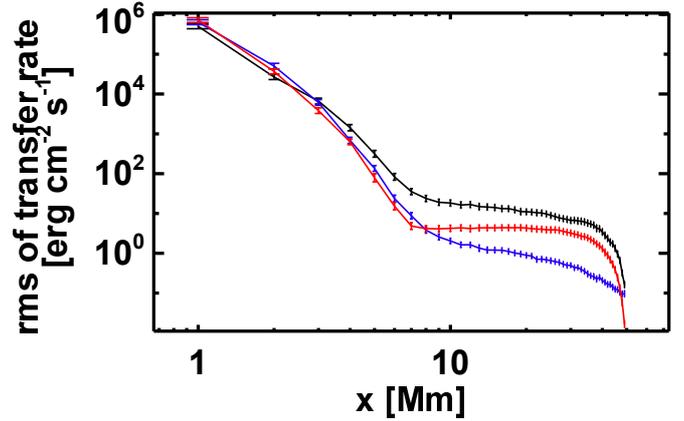} 
 \end{center}
\caption{
Root mean squares of energy transfer rates as functions of $x$.
Solid black, red, and blue lines represent $T_{\rm KBT}$, $T_{\rm KBA}$, and $T_{\rm KBC}$, respectively, and error bars for each point represent 95\% confidence intervals.
}\label{fig:transfer_log}
\end{figure}

The time evolution of the magnetic PSD at a certain horizontal wave number $k_\perp$ can be described as
\begin{eqnarray}
 \frac{d}{dt} E_{\rm M}(x,k_\perp) = T_{\rm KBT}(x,k_\perp) + T_{\rm KBA}(x,k_\perp) + T_{\rm KBC}(x,k_\perp). \label{eq:transfer}
\end{eqnarray}
The terms in Eq. \ref{eq:transfer} can be defined as
\begin{eqnarray}
 E_{\rm M}(x,k_\perp)  &\equiv& \frac{W^2}{4\upi^2}\oint \frac{1}{8\upi} 
 (\hat{\mathbfit{B}},\hat{\mathbfit{B}}) k_\perp d\theta, \\
 T_{\rm KBT}(x,k_\perp)  &\equiv& \frac{W^2}{4\upi^2} \oint \frac{1}{4\upi} 
 (\hat{\mathbfit{B}},\widehat{[\mathbfit{B}\cdot\nabla\mathbfit{v}]}) k_\perp d\theta, \\
 T_{\rm KBA}(x,k_\perp)  &\equiv& - \frac{W^2}{4\upi^2} \oint \frac{1}{4\upi} 
 (\hat{\mathbfit{B}},\widehat{[\mathbfit{v}\cdot \nabla\mathbfit{B}]}) k_\perp d\theta, \\
 T_{\rm KBC}(x,k_\perp)  &\equiv& - \frac{W^2}{4\upi^2} \oint \frac{1}{4\upi} 
 (\hat{\mathbfit{B}}, \widehat{[\mathbfit{B}\nabla \cdot\mathbfit{v}]}) k_\perp d\theta,
\end{eqnarray}
where
\begin{eqnarray}
 \hat{f} &=& \frac{1}{W^2} \int^{W/2}_{-W/2} f(x,y,z) e^{-i(k_y y + k_z z)} dy dz, \\
 (\mathbfit{A},\mathbfit{B}) &=& \frac{1}{2} (\mathbfit{A}\cdot\mathbfit{B}^* + \mathbfit{A}^*\cdot\mathbfit{B})
\end{eqnarray}
for an arbitrary function $f$, arbitrary vectors $\mathbfit{A}, \mathbfit{B}$, 
$k_\perp=\sqrt{k_y^2+k_z^2}$, and $\tan{\theta} = k_y/k_z$.
The transfer functions, $T_{\rm KBT},T_{\rm KBA},$ and $T_{\rm KBC}$ provide the energy transfer rate from the 
kinetic-energy reservoir to the $k_\perp$ component of the magnetic-energy reservoir 
via the magnetic tension, advection, and compressible terms 
in the induction equation, respectively \citep{2010ApJ...714.1606P}.
In Fig. \ref{fig:transfer_log}, the root mean squares of the transfer functions at $2\upi /k_\perp= $ 1 Mm are plotted as functions of $x$.
The solid black, blue, and red lines indicate $T_{\rm KBT}$, $T_{\rm KBA}$, and $T_{\rm KBC}$, respectively.
The temporal average was taken over $t\in[30, 140]$ min with a sampling time of 17 s such that 318 total sampling points were used to derive a 95-\% confidence interval for each height.
The contribution of the compressible term ($T_{\rm KBC}$) was non-negligible until slightly above the magnetic canopy ($x \lesssim 3$ Mm), while the tension term was dominant in the upper chromosphere and the corona.
The contribution of the compressible term around the magnetic canopy was comparable with that of the tension term up to $2\upi /k_\perp>0.4$ Mm in the current simulation.
At a smaller scale ($2\upi /k_\perp<0.4$ Mm), the advection term dominated the energy transfer.

\section{Discussion}
We performed 3D MHD simulations based on a single closed flux tube and found that the continuous perturbations provided to the foot points propagated upwards as MHD waves and deposited their wave energy, maintaining a 
1-MK temperature 
corona.
The mean temperature, pressure, and half-loop length were 0.9 MK, 0.9 erg cm$^{-3}$, and 45 Mm, respectively. The resulting loop could be considered over-dense compared with that predicted by the RTV scaling law.
Major heating events occurred sporadically in a highly localised region and were associated with collisions between counter-propagating Alfv\'{e}n waves.

Although our model was consistent with the observed cool loop in that the loop was either over-dense or over-pressure \citep{2001ApJ...550.1036A}, 
the velocity amplitude at the loop top, $\sqrt{v_{\rm y}^2+v_{\rm z}^2}$, of $\sim$ 60 km s$^{-1}$ in our model was significantly larger than those estimated from observations of $\sim$ 20 km s$^{-1}$ \citep{1999ApJ...513..969H,2016ApJ...820...63B}.
This discrepancy was also verified by the results of RMHD simulations \citep{2017ApJ...849...46V}, although the 
difference was smaller than that of our model.
Because the averaged density at the loop top in our model was $\sim10^8$ cm$^{-3}$, which was an order of magnitude less than 
that of the active region loop, the kinetic energy density between our results and the previous observations may not have differed drastically.


It was determined that the assumption in the RMHD approximation in the corona was valid, whereas the assumption in the chromosphere was not.
The RMHD approximation assumed small magnetic fluctuations, incompressibility, and a small flux tube radius
\citep[e.g.,][]{2011ApJ...736....3V}.
In our model, the amplitudes of the fluctuations were larger than the local Alfv\'{e}n speed near the magnetic canopy.
The radius of the flux tube below the canopy was comparable with the scale height of the field strength.
A measure of compressibility was provided by the sonic Mach number, whose value approached 1 in the chromosphere.
Moreover, the transfer function of the compressible term was comparable with those of the tension and the advection terms.
Therefore, in addition to the incompressible processes, the compressible processes are important, at least 
at the large scale ($2\upi /k_\perp>0.4$ Mm) near the magnetic canopy.

The resulting model atmosphere is consistent with the open field regions above the polar region rather than with the active region loops.
The height of the magnetic canopy in our model was somewhat consistent with that predicted theoretically by
\cite{1976RSPTA.281..339G}.
However, the 5-Mm height of the transition region in our model was significantly larger than the standard value of 1.7-2.3 Mm \citep{1993ApJ...406..319F}.
The high chromospheric temperature ($\sim 7,000$K) or the high pressure scale height, 
which are highly sensitive to the treatment of the chromospheric cooling, could have caused the difference.
Based on EUV observations, \cite{1998ApJ...504L.127Z} suggested that the height of the transition region was systematically 6 Mm higher in the polar region than at the equator.
Taller spicules observed in the polar region \citep{2012ApJ...750...16Z,2012ApJ...759...18P} may be 
a signature of a higher transition region \citep{1982SoPh...78..333S}.
In addition, lower coronal temperatures in the coronal hole also result in taller spicules and higher transition regions \citep{2015ApJ...812L..30I}.

Our model demonstrated that nanoflare-like heating, in the sense that the heating processes were
intermittent and localised, was possible, even in the wave heating mechanisms.
The primary difference between the stressing and wave models may be the ratio between the kinetic and the magnetic energy in the fluctuation.
In the cases of the wave models, the kinetic and magnetic energy are comparable, while the magnetic energy is dominant in the stressing models \citep{2008ApJ...677.1348R}.

The average and maximum temperatures obtained in our model would increase when the spatial resolution improves. Because the MHD waves in the chromosphere dissipate numerically, the amount of wave energy that is 
able to reach the corona increases along with the resolution, resulting in a higher heating rate in the corona.
This phenomenon was observed in previous 2D MHD simulations \citep{2016MNRAS.463..502M}.

In our model, plausible heating mechanisms, such as phase mixing \citep{1983A&A...117..220H} 
and resonant absorption \citep{1978ApJ...226..650I}, were artificially suppressed by our assumption of periodic boundary.
Because of this assumption, the system was filled with identical flux tubes, and the isolated flux tubes typically seen outside of the active regions were not realised in our model.
This assumption suppressed the density jump between inside and outside of the flux tube that is essential for the phase mixing and the resonant absorption.
However, it is possible that, even under our model, the phase mixing and resonant absorption will remain effective by using the density inhomogeneity inside the flux tube. Further analysis is needed to distinguish among the different wave heating mechanisms.

Realistic simulations that utilize either radiative transfer equations or precise equation of states 
\citep{2015ApJ...811..106H,2017ApJ...834...10R} are important.
The present study relied on an empirical cooling rate that resulted in temperature errors of a few thousand degree Kelvin
in the chromosphere. 
Such errors can directly affect the dynamics of compressible waves.
In addition, the dynamics of incompressible waves can be affected indirectly via either mode coupling with compressible waves or atmospheric structures, such as density scale height and the height of the transition region.

\section{Conclusions}
The coronal heating problem is one of the most difficult and longest-studied problems in solar physics, having remained unsolved for more than half century.
Although numerous studies have been devoted to wave heating mechanisms, thermal responses to the dynamic wave dissipation process have not yet been determined.
To investigate the relationships between thermal properties and wave dynamics, 
we performed 3D MHD simulations for a single flux tube. Our main conclusions were:

\begin{enumerate}
 \item A 
1-MK temperature 
corona was reproduced via wave propagation and dissipation when fluctuations were continuously applied to the foot point of the flux tube. The mean temperature, pressure, and half-loop length were 0.9 MK, 0.9 erg cm$^{-3}$ s$^{-1}$, and 45 Mm, which represented an over-dense, over-pressure loop in comparison to that assumed by RTV theory.
 \item The high-temperature region was created intermittently in our simulation and was highly localised.
       In addition, the wave heating can be considered nanoflare heating in terms of intermittency and locality.
 \item The velocity dispersion obtained in the corona ($\sim$60 km s$^{-1}$) was significantly larger than those
       estimated from observations ($\sim$ 20 km s$^{-1}$).
       This demonstrates that the current model cannot be directly applied to the active region loops.
 \item The RMHD approximation assumptions were well satisfied in the corona.
       However, this was not the case around the magnetic canopy, where the assumptions were violated and where the compressible terms played an important role in energy transfer perpendicular to the background magnetic field.
\end{enumerate}

Although our simulations revealed several new properties, the results of only a single set of parameters were provided.
To fully understand the thermal responses of the corona to photospheric motion, the large set of parameters consisting of
the properties of the foot point forces, structures of the magnetic field, and system size has to be surveyed.
The application of our model to the open field region is another topic of interest for investigation of solar wind acceleration problems \citep{2005ApJ...632L..49S,2006JGRA..11106101S,2012ApJ...749....8M,2014MNRAS.440..971M}.

\section*{Acknowledgements}

The author appreciates the anonymous referee for providing constructive comments.
The numerical computations were conducted on a Cray XC30 supercomputer at the Centre for Computational Astrophysics,
National Astronomical Observatory of Japan.
A part of this study was performed by using the computational resource of the Center for Integrated Data Science,
Institute for Space-Earth Environmental Research, Nagoya University through the joint research program.
This work was supported by JSPS KAKENHI, Grant Number 50728326.

\hyphenation{Post-Script Sprin-ger}








\bsp	
\label{lastpage}
\end{document}